\documentclass[aps,prl,twocolumn,superscriptaddress,notitlepage,nofootinbib]{revtex4-1}
\usepackage{graphicx}
\usepackage{epstopdf}
\usepackage{bm}
\usepackage{textcomp}
\usepackage{epsfig}
\usepackage{graphics}
\usepackage{xspace}
\usepackage{amssymb}
\usepackage{amsmath}
\usepackage{xcolor}
\usepackage[utf8]{inputenc}
\usepackage[normalem]{ulem}
\usepackage{stmaryrd}
\usepackage[inline]{enumitem}
\def\kslash{k\!\!\!\slash}
\def\bnslash{\bar n\!\!\!\slash}

\def\OMIT#1{}

\newcommand{\nn}{\nonumber}

\newcommand{\bea}{\begin{eqnarray}}
\newcommand{\eea}{\end{eqnarray}}

\newcommand{\gsim}{\mathrel{\rlap{\lower4pt\hbox{\hskip1pt$\sim$}}\raise1pt\hbox{$>$}}}

\newcommand{\Pythiaeight}{\textsc{Pythia}8\xspace}

\newcommand{\be}{\begin{equation}}
\newcommand{\ee}{\end{equation}}

\begin{document}
\title{The Time-reversal Odd Side of a Jet}
 \author{Xiaohui Liu}
\email{xiliu@bnu.edu.cn}
\affiliation{Center of Advanced Quantum Studies, Department of Physics, Beijing Normal University, Beijing 100875, China}
\affiliation{Center for High Energy Physics, Peking University, Beijing 100871, China}
\author{Hongxi Xing}
\email{hxing@m.scnu.edu.cn }
\affiliation{Guangdong Provincial Key Laboratory of Nuclear Science, Institute of Quantum Matter, South China Normal University, Guangzhou 510006, China}
\affiliation{Guangdong-Hong Kong Joint Laboratory of Quantum Matter,
South China Normal University, Guangzhou 510006, China}

\begin{abstract}
We re-examine the jet probes of the nucleon spin and flavor structures. We find for the first time that the time-reversal odd (T-odd) component of a jet, conventionally thought to vanish, can survive due to the non-perturbative fragmentation and hadronization effects. This additional contribution of a jet will lead to novel jet phenomena relevant for unlocking the access to several spin structures of the nucleon, which were thought to be impossible by using jets. As examples, we show how the T-odd constituent can couple to the proton transversity at the Electron Ion Collider (EIC) and can give rise to the anisotropy in the jet production in $e^+e^-$ annihilations. We expect the T-odd contribution of the jet to have broad 
applications in high energy nuclear physics. 
\end{abstract}

\maketitle

{\it Introduction.} 
%Jet, very limited spin structure probes. T-odd never discussed.  
With the advent of the Electron Ion Colliders,  EIC~\cite{AbdulKhalek:2021gbh} and EicC~\cite{Anderle:2021wcy}, the studies of the femtoscale structure of the nucleon are entering a new era, which will portray the full three-dimensional (3D) image of the nucleon with unprecedented precisions. To effectively decipher the spin and flavor information from the experimental data, one requires the global analyses of the unpolarized and polarized parton distribution functions (PDFs) and their 3D counterparts such as the transverse momentum distributions (TMDs). The global analyses will simultaneously utilize all possible hard probes including inclusive cross sections, semi-inclusive hadron productions and jets, from all available programs such as the COMPASS, HERMES and the Relativistic Heavy Ion Collider (RHIC). Each probe has its own advantage and complements to the others.

The idea of the jets, which originally emerged as a primary technique for studying the strong interactions, 
now further demonstrates its power in probing the fundamental properties of the nucleon and the nuclear medium through the measurements at the LHC and RHIC~\cite{Aschenauer:2016our,Adamczyk:2017wld,Boer:2014lka,Connors:2017ptx,Cao:2020wlm}. The high luminosity of the future EIC considerably boosts the studies of the jets and the jet substructures 
for exploring the nucleon spin structures and great progress has been made. The jets have been showed to have the ability to access the unpolarized TMDs as well as the transversely polarized Sivers functions~\cite{Kang:2011jw, Liu:2018trl,Arratia:2020nxw,Liu:2020dct,Aschenauer:2019uex,Arratia:2019vju,Arratia:2020ssx,Kang:2020xyq,Aschenauer:2017jsk,Zheng:2018ssm,Hatta:2020bgy}. The jet substructure studies, like the soft-drop grooming~\cite{Larkoski:2014wba}, hadron in jets~\cite{Kang:2017glf,Kang:2017btw},  the perturbative interference effect~\cite{Chen:2020adz}
as well as newly developed jet recombination and construction schemes~\cite{Gutierrez-Reyes:2018qez, Gutierrez-Reyes:2019vbx,Arratia:2020ssx},  such as the ``winner-take-all" (WTA) scheme~\cite{Gutierrez-Reyes:2018qez, Gutierrez-Reyes:2019vbx}, further enrich the content of the jet probes.

However, compared with the conventional semi-inclusive deep inelastic scattering (DIS) process, the jet probe of the nucleon structures faces two fundamental difficulties: 
\begin{enumerate*}
\item It is hard to achieve the quark flavor separation. 
\item It only couples to extremely limited nucleon spin structures, due to the unpolarized and time-reversal even nature of a jet. 
\end{enumerate*}
Recently proposed jet charge resolves the first issue~\cite{Kang:2020fka}, while the second remains challenging and limits the power of the jet probe. Specifically,
all the time-reversal odd (T-odd) proton distributions, such as the transversity which is crucial for extracting the least known nucleon tensor charge~\cite{Ralston:1979ys, Jaffe:1991kp, Jaffe:1991ra, Cortes:1991ja, Gamberg:2001qc}, are regarded as unreachable directly by the jets. See~\cite{Kang:2020xyq} for recent efforts.
\begin{figure}[htbp]
\vspace{-5.ex}
  \centering
\includegraphics[width=.98\linewidth]{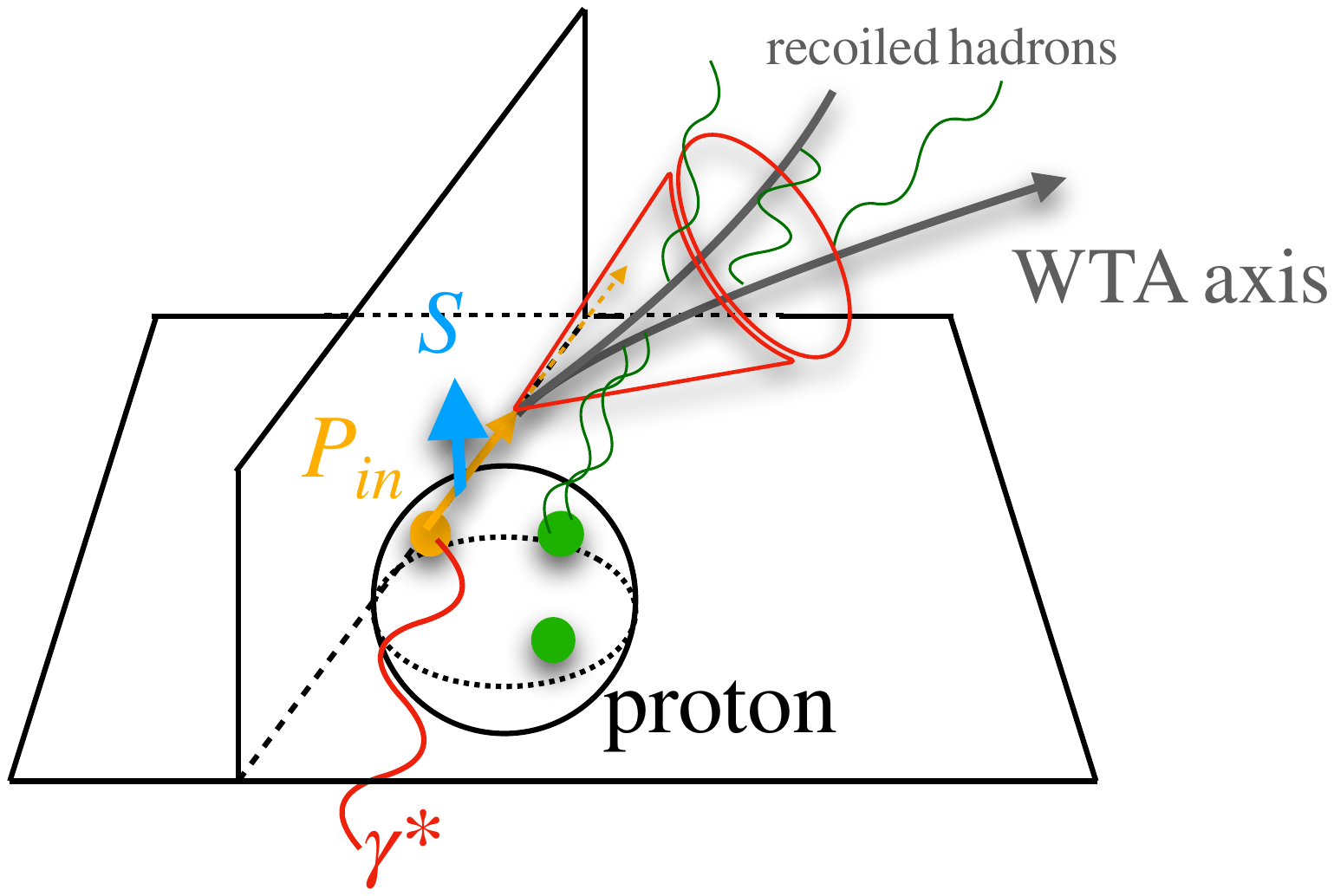} 
 \vspace{-5.ex}
\caption{Origin of the jet T-odd contributions. \OMIT{The WTA jet axis lies outside the plane by the spin $S$ and $P_{in}$, to allow for the asymmetry due to the quantum correlation between parton’s spin and its hadronization about the plane.}}
% \vspace{1.ex}
\label{fig:wta-axis}
\end{figure}

In this work, contrary to the conventional wisdom, we notice that the jets can acquire the T-odd contributions due to its non-perturbative ingredients,  especially in the ${\cal O}(\sim  1\, {\rm GeV})$ region where the interesting nuclear Femtography is performed~\cite{qiutalk}. 
The basic idea roots in the fact that the final state interactions induce an asymmetry for the partons with different spin orientations $S$~\cite{Brodsky:2002cx, Ji:2002aa,Amrath:2005gv} that initiate the jets. Similar to the Collins effect~\cite{Collins:1992kk} in semi-inclusive DIS, the asymmetry can be observed as long as the jet axis is not aligned with the fragmenting parton momentum $P_{in}$. 
%the spin direction $S$ of the fragmenting parton generates a preference in its hadronization or an asymmetry, in one side of the plane determined by the parton momentum  $P_{in}$ and its spin direction $S$, than the other side. Similar to the Collins effect~\cite{Collins:1992kk}, as long as the jet axis is not aligned with the fragmenting parton, this asymmetry will be non-vanishing due to the nontrivial phases induced by non-perturbative final state interactions. 
This is the generic case for the WTA jets as sketched in fig.~\ref{fig:wta-axis} as well as the groomed jets\OMIT{, where the jet axis does not fall into the plane spanned by $P_{in}$ and $S$}. The situation also holds for the standard jets due to the soft radiations outside the jet cone~\cite{Cal:2019gxa}.  
The small imbalance between the jet axis and the parton momentum direction in the  $\sim {\cal O}(1\, {\rm GeV})$ region\OMIT{, where roots the focus of most nucleon spin programs,} could thus probe the asymmetry. 
\OMIT{One can further enhance the net asymmetry by measuring the jet charge to separate the fragmenting parton species~\cite{Kang:2020fka}. Therefore, we expect this non-perturbative T-odd side of a jet to be feasible by the on-going and future spin programs.}

The T-odd jet %\textcolor{red}{is universal and}
couples directly to the T-odd proton distributions, therefore immediately opens up many unique opportunities for probing the nucleon intrinsic spin dynamics using jets, which were thought to be impossible or difficult. As concrete examples, we demonstrate the feasibility and advantage of T-odd jets to the proton transversity in transversely polarized DIS by extending the factorization of the conventional unpolarized jet production to the T-odd case. We also show the same T-odd jets can be constrained directly through the azimuthal angle asymmetry in the $e^+e^-$ annihilation.

{\it QCD factorization and the Time-reversal odd jet function.} To introduce the T-odd jet function, we consider jet production in the collision between an unpolarized electron $e$ and a transversely polarized proton $p$ with spin $s_T$ and momentum $P$,  $l\, p(P, s_T)  \to l' \, J(P_J) + X$, % in the Breit frame where
%the four momentum of the virtual photon $q^\mu = (0,0,0,-Q)$ and 
 where we assume $P^\mu = \frac{Q}{2}(1,0,0,1) = \frac{Q}{2} n^\mu$. 
We construct the jet out of the WTA scheme~\cite{Gutierrez-Reyes:2018qez,Gutierrez-Reyes:2019vbx} with radius $R$. We emphasize that the use of the WTA jet is for illustration. Our conclusion holds generically for jet production. We are particularly interested in the case that $q_T \ll Q$, especially for $q_T \sim \Lambda_{\rm QCD}$ to probe the intrinsic dynamics of the proton. Here $q_T = q- \frac{n\cdot q}{2} {\bar n} - \frac{{\bar n}\cdot q}{2} n$ is the transverse momentum of the virtual photon in the frame that the jet axis defines the conjugate direction ${\bar n} = (1,0,0,-1)$. Here $q$ denotes the virtual photon momentum.  In the frame that the virtual photon and the proton develop no transverse momenta (the $\gamma$-$p$ frame), $q_T = - \frac{P_{J\perp}}{z}$ up to $1/Q^2$ corrections where $z = n\cdot P_J/Q = P_J^-/Q$. Here we introduce the notation that for an arbitrary vector $v$, $n\cdot v \equiv v^- $ while ${\bar n} \cdot v  \equiv v^+$. 

We follow~\cite{Gutierrez-Reyes:2018qez,Gutierrez-Reyes:2019vbx} to derive the factorization theorem for the cross section using Soft Collinear Effective Theory (SCET)~\cite{Bauer:2000yr, Bauer:2001ct, Bauer:2001yt, Bauer:2002nz}. In the small $q_T$ limit, the $n$ and ${\bar n}$ collinear sectors can be factorized and the cross section can be written as~\cite{Gutierrez-Reyes:2018qez} 
\bea
&& \sigma =  \frac{1}{2 s }
\int [dl'] \frac{e^4e_q^2}{Q^4}  \frac{1}{2} L_{\mu\nu}
% \nn \\
%&\times& 
 \frac{1}{2 N_c} \,
\int   d^2 x_T \, e^{i q_T \cdot x_T}
 \gamma^\mu_{\alpha\beta}  \gamma^\nu_{\beta'\alpha'}
 \nn \\
& \times &
 \int d x^-
 e^{i q^+  x^-}
\langle P, s_T|
\xi_n (0)
\xi^\dagger_{n}(x^-,x_T) 
 | P, s_T\rangle_{\alpha'\alpha}  \nn \\
&\times & \int [dP_J] 
 d x^+ 
e^{i q^-  x^+}  %\nn \\
%&&\times  
\langle 0 |
 \xi_{\bar n} (x^+,x_T)
% | J(Q_{J_h}) X_{\bar n} \rangle_{\beta} 
% \langle J(Q_{J_h}) X_{\bar n} | 
 | J  X_{\bar n} \rangle_{\beta} 
 \langle J X_{\bar n} | 
 \xi^\dagger_{\bar n}(0) |0\rangle_{\beta'} \,, 
\,  \nn \\
  \eea 
  where $s$ is the center of mass energy squared and $L^{\mu\nu}$ is the leptonic tensor. The notation $[d \dots ]$ denotes the phase space measure and $\xi_{n(\bar n)} = W_{n({\bar n})}^\dagger\chi_{n({\bar n})} $~\cite{Bauer:2000yr, Bauer:2001ct, Bauer:2001yt, Bauer:2002nz} is the SCET quark field encodes both the $n({\bar n})$-collinear quark field $\chi_{n({\bar n})}$
 and the Wilson line $W_{n({\bar n})}$. 
 Here 
 $J$ reminds us that a signal jet is constructed in the final state.
 
 To proceed, one introduces the proton structure function $\Phi_{\alpha'\alpha}$
and the jet function matrices ${\cal J}^q_{\beta\beta'}$, such that 
\bea
&& \langle P, s_T|  {\bar \xi}_{n}(0) {\xi}_n(x^-, x_T) |P,s_T \rangle_{ \alpha'\alpha}  \nn \\
&=& P^+ \int_0^1 d \zeta  \int d^2 p_T 
e^{i \zeta \, P^{+}  x^- }
e^{i p_T \cdot x_T } \Phi_{\alpha'\alpha}(\zeta,p_T) \,, 
\eea
and 
\bea\label{eq:jet-tensor}
&& 
\langle 0 |   {\bar \xi}_{{\bar n}}(x^+,x_T) | J X_{\bar n} \rangle_{\beta} 
 \langle J  X_{\bar n} | {\xi}_{\bar n} (0)   |0 \rangle_{\beta'}   \nn \\
&=& 2P_J^- 
\int_0^1 \frac{d z}{z} \int d^2 k_T 
e^{-i  \frac{ P_J^{-}}{z}  x^+ }
e^{-i k_T \cdot x_T } {\cal J}^q_{\beta\beta'}(z,k_T,R ) \,,\, \quad %\nn\\
\eea
to find the cross section in the form:  
 \bea\label{eq:factorization}
&& \sigma =  \frac{1}{2 s }
\int [d l'] \frac{e^4e_q^2}{Q^4}  \frac{1}{2} L_{\mu\nu}
 \nn \\
&& \times 
%\frac{z}{2 N_c} \,2\,
\frac{1}{2 N_c} \,2\,
\int [d P_J] \,  
  \int d^2 p_T \int d^2 k_T (2\pi)^4 \delta^{(2)}( q_T+ p_T-k_T)
   \nn \\
&&  \times \delta(1-z)
\gamma^\mu_{\alpha\beta}  \gamma^\nu_{\beta'\alpha'} \,
 {\mathfrak J}^q_{\beta\beta'}(k_T) \,  \Phi_{\alpha'\alpha}(\zeta,p_T) \,,
  \eea 
where $\zeta = Q/P^+$ and $z = P_J^-/Q$. Here we have used the property that for the WTA jet, when $k_T \ll P_J R $, $ {\cal J}^q_{\beta\beta'}(z,k_T,R) = \delta(1-z) {\mathfrak J}^q_{\beta\beta'}(k_T) + {\cal O}\left( \frac{k_T^2}{p_J^2R^2} \right)$, independent of both $R$ and $z$~\cite{Gutierrez-Reyes:2018qez}. \OMIT{\textcolor{red}{We emphasize that the use of the WTA jet is for illustration but not mandatory. Our conclusion holds generically for jet production.}}

According to spin structures, 
the matrix element ${\mathfrak J}^q_{\beta\beta'}$ can take the general form that  (see for instance~\cite{Bauer:2002nz,Bacchetta:2006tn})
\bea\label{eq:jet}
{\mathfrak J}^q(k_T)
= \frac{\bnslash}{2}\, J^q(k_T)
+i \frac{\kslash_T \bnslash}{2 } J^q_{T}(k_T) \,, 
\eea
where the first term $J^q(k_T)$
is the standard WTA jet function which has been discussed extensively in the literature. 

Conventionally, the second term $J^q_{T}(k_T)$ is always discarded as it is T-odd and therefore regarded as
vanishing due to the time-reversal symmetry of perturbative QCD. 
Therefore, only the first term in Eq. (\ref{eq:jet}) survives in all available theoretical and experimental discussions, which eventually leads to the limited power of the jet as the probe of nucleon spin structures. For instance, using jets to probe the nucleon transversity and the Boer-Mulders function was considered to be impossible.

However, \OMIT{in the scenario we considered and also}in most spin programs, since $k_T \sim \Lambda_{\rm QCD}$, the jets are sensitive to the strong interaction fragmentation and hadronization. Typical examples are the TMD studies using jets proposed in~\cite{Gutierrez-Reyes:2018qez,Gutierrez-Reyes:2019vbx} and~\cite{Arratia:2020ssx}. The relevant region is $q_T < 3 {\rm GeV}$ where the jets receive substantial non-perturbative contributions, which show up as the non-perturbative Sudakov factor of the jet function~\cite{Gutierrez-Reyes:2019vbx} or the shape function~\cite{Arratia:2020ssx} to be determined by fitting to the experimental data. Such non-perturbative dynamics 
can therefore induce the nontrivial phases from final state interactions and eventually prevents the T-odd jet function $J^q_{T}$ from vanishing,
%Exactly the same mechanism results in the Collins effect~\cite{Collins:1992kk} in the single parton fragmentation and 
%We expect the  consequence happens to the jets, 
especially for jets at machines with relatively low energy such as the 12 GeV CEBAF at Jefferson Lab, EIC, EicC and etc..

The factorization theorem guarantees the 
universality and the predictive power of the conventional jet function $J^q$, which also holds true for the T-odd jet function:
\begin{itemize}
\item From Eq.~(\ref{eq:jet-tensor}) and Eq.~(\ref{eq:jet}), it immediately follows that the T-odd jet function $J_T^q$ is universal  as its conventional counterpart $J^q$. 
\item Similar to the conventional jet studies~\cite{Becher:2013iya}, in the Fourier transformation position $b$-space, the T-odd jet $J_T^q$ can be written as a product of the perturbative calculable coefficient and the non-perturbative normalization by the operator studies following~\cite{Yuan:2009dw} and~\cite{Becher:2013iya}. The jet algorithm dependence will be included in the perturbative coefficient. As a consequence, the T-odd $J_T^q$ shares the comparable predictive power as $J^q$ in the small $k_T$ region relevant for the nucleon TMD and spin studies. It will be particularly interesting to note  for the WTA jet, that the only non-perturbative contribution to the $J_q$ is the non-perturbative Sudakov and therefore can be fully determined by lattice calculations~\cite{Shanahan:2020zxr}. We expect the similar story will happen to the $J^q_T$, that the non-perturbative normalization may be calculable by available techniques~\cite{Shanahan:2020zxr, Zhang:2020dbb}. However this can not be done for the TMD fragmentation functions by any known techniques due to the explicit final state hadron tagged.  
\end{itemize}
Besides, the flexibility of choosing the jet recombination schemes and hence the jet axis, allows us to adjust the sensitivity to different non-perturbative contributions. This extra control could provide the opportunity to ``film" the QCD non-perturbative dynamics, if one manages to continuously change the axis from one to another. 

The T-odd jets\OMIT{function {could} \textcolor{red}{and the control of jet recombination schemes will}} open up many new opportunities for probing the proton intrinsic dynamics, which were thought to be unreachable by using jets. Below we highlight some of the possible applications to the nucleon structure studies. 

{\it Transversely polarized deep-inelastic-scattering.} From Eq.~(\ref{eq:factorization}) and Eq.~(\ref{eq:jet}) and compare with the hadron production~\cite{Bacchetta:2006tn}, one can immediately realize that the non-vanishing T-odd component allows the jet to probe the proton T-odd distributions. For example, the nucleon transversity $h_{1}$ can be accessed through the azimuthal distribution 
 \bea\label{eq:dis-jet}
%&&
A(\zeta,y,\phi_s,\phi_J,P_{J\perp})
%\frac{d\sigma^{\sin(\phi_J+\phi_s)}_{UT}}{d\zeta dy\,d\psi \,d\phi_J dP_{J\perp}^2} 
%\nn \\ 
= % \frac{\alpha^2}{\zeta \, y Q^2} \frac{y^2}{2(1-\epsilon)}
%\left(1+\frac{\gamma^2}{2\zeta} \right)
1+ 
 \epsilon  |s_\perp| 
\sin(\phi_J+\phi_s) % \nn \\
%&\times&  
\frac{F_{UT}}{F_{UU}} \,,
%\sum_q\, e_q^2 \, 
%\int \frac{d^2 b}{4\pi^2} e^{-i  P_{J\perp} \cdot b} 
%%%e^{-S(b,Q)-S_{NP}(b,Q/Q_0)}
%%%\,  \nn \\
%%%&\times&  %\int d^2p_T d^2 k_T  \,
%%%e^{i(p_T-k_T)\cdot b_T}  
%%%\delta\left( \frac{-P_{J\perp}}{z}+ p_T- k_T\right) \nn\\
%i  \frac{  P_{J\perp}^\alpha }{P_{J\perp}}\, 
%  \, \zeta  \, h^q_{1T}(\zeta,b)\,
% \partial_{b^{\alpha}}
% J^q_{T}(b,Q_{J_h})\,. \nn \\
\eea
where $F_{UU}$ is the conventional unpolarized jet production which can be found in~\cite{Gutierrez-Reyes:2018qez} and the azimuthal dependent part
\bea\label{eq:FUT}
F_{UT} = \sum_q\, e_q^2 \, 
\int \frac{d^2 b}{4\pi^2} e^{-i  P_{J\perp} \cdot b} 
%e^{-S(b,Q)-S_{NP}(b,Q/Q_0)}
%\,  \nn \\
%&\times&  %\int d^2p_T d^2 k_T  \,
%e^{i(p_T-k_T)\cdot b_T}  
%\delta\left( \frac{-P_{J\perp}}{z}+ p_T- k_T\right) \nn\\
i  \frac{  P_{J\perp}^\alpha }{P_{J\perp}}\, 
  \, \zeta  \, h^q_{1}(\zeta,b)\,
 \partial_{b^{\alpha}}
 J^q_{T}(b)\,. \nn \\
\eea
%$\hat{h} = P_{J\perp}/|P_{J\perp}|$. 
Here $h^q_{1}(\zeta,b)$ and $J^q_T(b)$ are the Fourier transformation of the proton transversity $h^q_{1}(\zeta,p_T)$ and  the T-odd jet function $J^q_T(k_T)$, respectively.  

\begin{figure}[htbp]
\vspace{-7.ex}
  \centering
\includegraphics[width=1\linewidth]{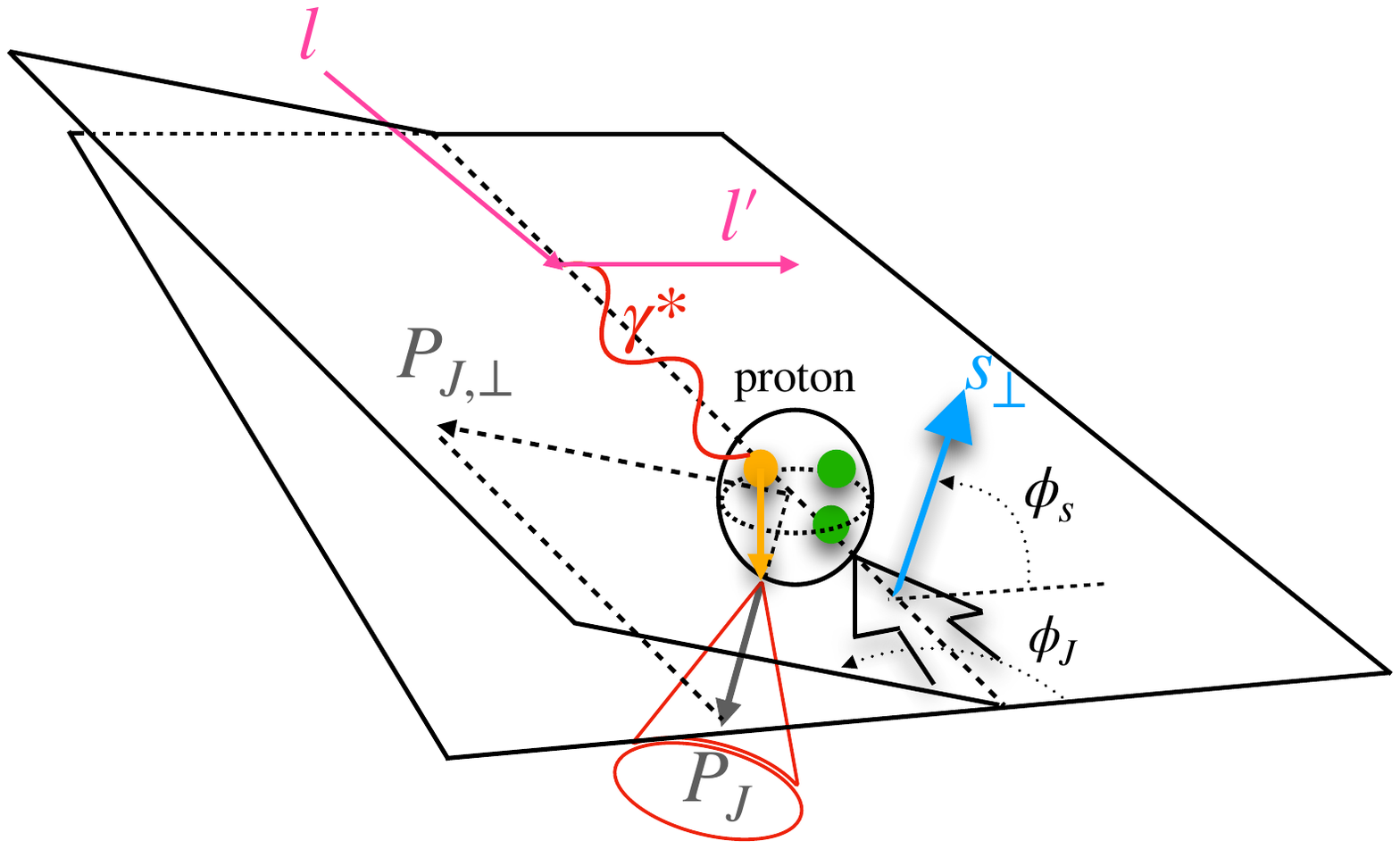} 
 \vspace{-7.ex}
\caption{Angular variables involved in Eq.~(\ref{eq:dis-jet}) in DIS.}
 \vspace{1.ex}
\label{fig:ep-frame}
\end{figure}

The azimuthal distribution is measured in the $\gamma$-$p$ frame, where the kinematics involved are illustrated in Fig.~\ref{fig:ep-frame} 
 with
\bea
\epsilon = \frac{1-y-\frac{1}{4}\gamma^2 y^2}{1-y+\frac{1}{2}y^2+\frac{1}{4}\gamma^2y^2}\,, \quad \quad
y = \frac{P\cdot q}{P\cdot l }  \,,
\eea
and $\gamma =  2 \zeta \frac{m_P}{Q} $ with $m_P$ the proton mass.

Now we see clearly that the transversity can be measured via the jet probe which was thought to be impossible. Compared with measuring the final state hadron Collins effect, we see from Eq.~(\ref{eq:FUT}) that the T-odd WTA jet probe of the incoming proton transversity reduces the non-perturbative degrees of freedom, which allows for a cleaner extraction of the transversity at the future EIC and EicC. 

To access to this cross section, we construct jet using the WTA scheme and perform the measurements in the $\gamma$-$p$ frame. We restrict us to the region where $P_{J\perp} \sim \Lambda_{\rm QCD} \ll Q$. 
 \begin{figure}[htbp]
  \centering
\includegraphics[width=1\linewidth]{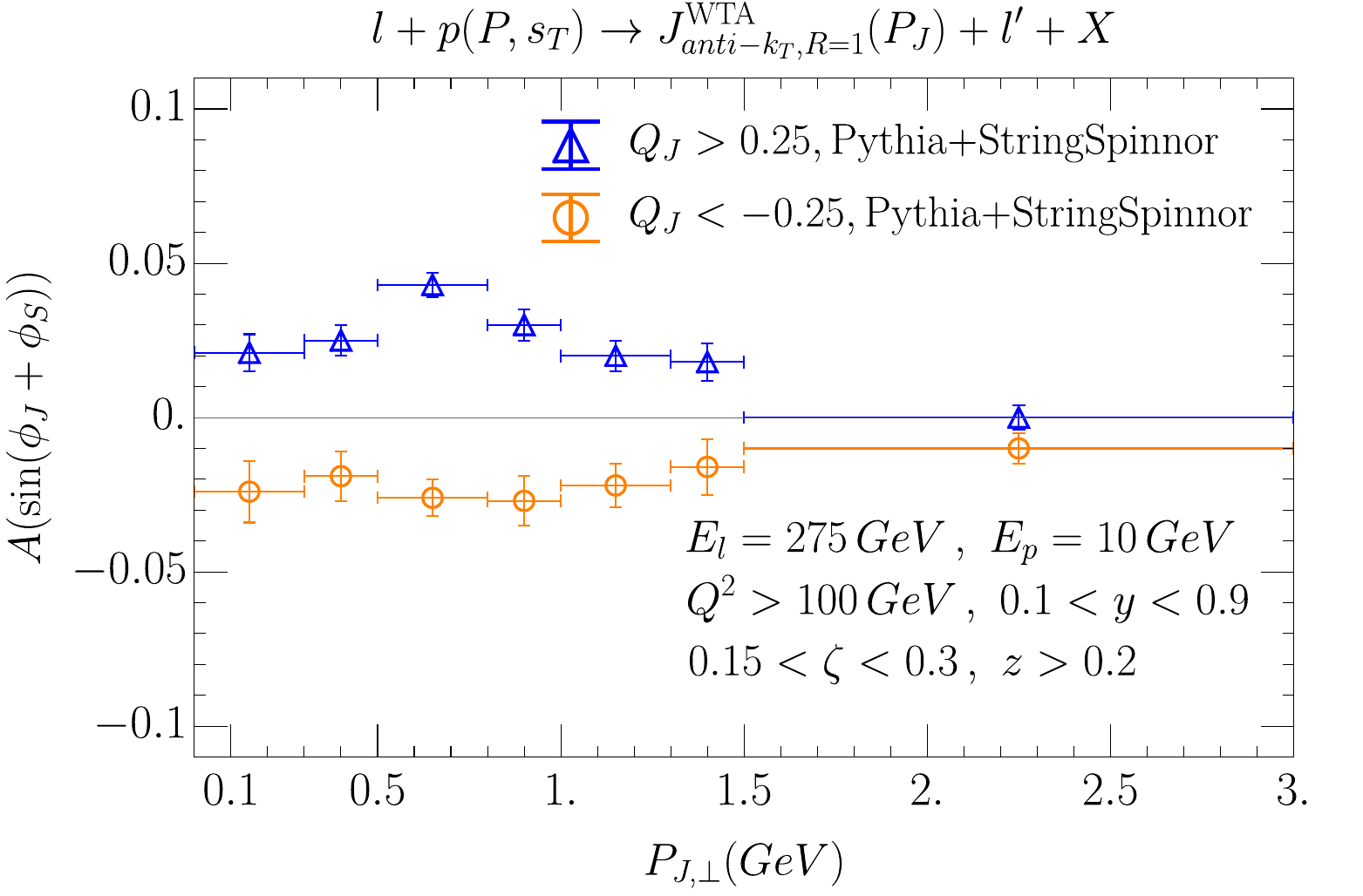} 
\caption{Azimuthal asymmetry by the proton transversity and the T-odd jet. Errors represent the statistic uncertainties. }
 \vspace{-3.ex}
\label{fig:Asymmetry-eic}
\end{figure}
One can avoid the possible cancellation between different quark flavors to further enhance the asymmetry by measuring the jet charge $Q_J$~\cite{Kang:2020fka}. The brief idea is to reweigh the relative contributions of different quark flavors to the jet by selecting the appropriate jet charge region.  For instance, it is found that a fraction of $r^+_{u ({\bar d})}=52\%$ of the $u\, ({\bar d})$ quark initiating jets will contribute to 
$Q_J > 0.25$ while only $r^+_{d ({\bar u})} = 15\%$ of the $d \, (\bar {u})$-quark jets contribute to this jet charge region~\cite{Kang:2020fka}. For $Q_J < -0.25$, the fractions change to $r_{u({\bar d})}^- = 15\%$
and $r_{d({\bar u})}^- = 52\%$.
Therefore the $d \, ({\bar u})$ quark contribution will be dramatically suppressed when $Q_J > 0.25$ and enhanced in $Q_J < - 0.25$.

A simulation of the azimuthal distribution generated by the transversity and the T-odd jet at the EIC is shown in fig.~\ref{fig:Asymmetry-eic} using \Pythiaeight~\cite{Sjostrand:2007gs} with the \textsc{StringSpinnor}\xspace plugin~\cite{Kerbizi:2021pzn}. The events are generated in the lab frame while the asymmetry is measured in the $\gamma$-$p$ frame. The simulation manifests the testable effects of the T-odd jets. Details on this simulation will be given in future publications.

 {\it Anisotropy in $e^+\, e^-$ annihilation.} 
 The same T-odd jet function will also arise in the back-to-back di-jet production in $e^+e^-$ annihilation, where the T-odd jets generate the angular anisotropy. Such asymmetry allows a direct and clean study of the T-odd jet via analyses of the BaBar or BELLE data.
 \begin{figure}[b!]
\vspace{-8.ex}
  \centering
\includegraphics[width=1\linewidth]{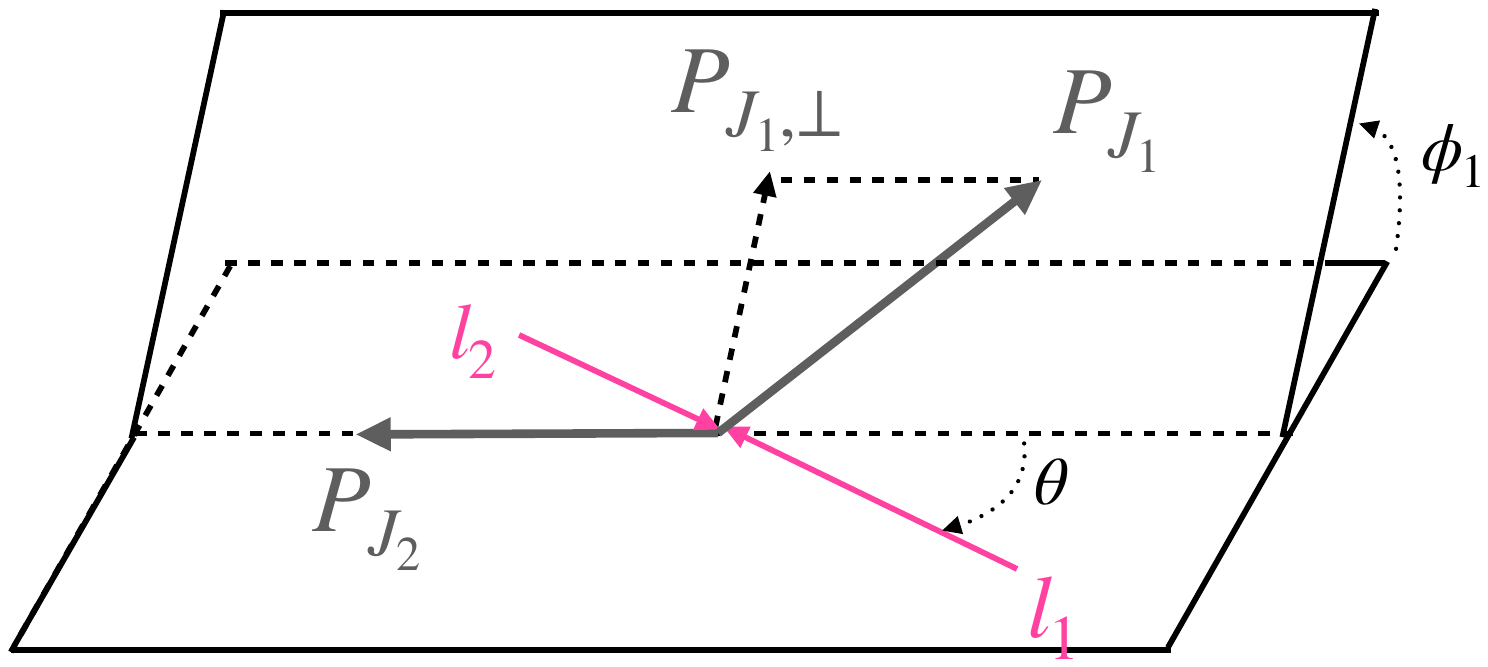} 
 \vspace{-12.ex}
\caption{Kinematics in the $e^+e^-$ annihilation.}
 \vspace{1.ex}
\label{fig:ee}
\end{figure}
Here again, we stick to the WTA jets with $q_T \ll \sqrt{s} R$, where $\sqrt{s}$ is the center of mass energy and %we have defined ${\hat h} = q_T/|q_T|$ with 
$q_T = - P_{J_1,\perp}$. The kinematics that we considered are depicted in Fig.~\ref{fig:ee}.

 The azimuthal asymmetry takes the general form~\cite{Kang:2014zza} 
 \bea
 R^{J_1J_2} = 
 1+ \cos(2\phi_1) \frac{\sin^2\theta}{1+\cos^2\theta} \frac{F_T(q_T)}{F_U(q_T)}\,,
 \eea
 where $F_U$ is the contribution from conventional jet, while $F_T$ from the T-odd component, respectively. The $F_U$ is known and found to be~\cite{Gutierrez-Reyes:2018qez}
 \bea
 F_U = q_T\, \sum_q e_q^2\, \int \frac{\mathrm{d}b \, b}{2\pi} J_0(q_T b) J^q(b){\bar J}^{\bar q}(b) \,, 
 \eea
 where $J_0(q_T b)$ is the Bessel function. The newly identified $F_T$ is given by
 \bea
  F_T &=& q_T \, \sum_q \, e_q^2\, \int \frac{\mathrm{d}^2b}{(2\pi)^2} e^{-iq_T \cdot b} 
  %\frac{
  \left( 2 \frac{q_T^\alpha}{q_T} \frac{q_T^\beta}{q_T} 
+ g^{\alpha\beta}  \right)
%}%{m_{h_1} m_{h_2}} 
\nn \\
&& \times \partial_{b^\alpha} J^q_{T}(b) \partial_{b^\beta}{\bar J}^{\bar q}_{T}(b) 
  \,.
 \eea
Here in $F_U$, we have
\bea
J^q(b){\bar J}^q(b) = e^{- S_{pert.} - S^J_{NP}} \left( 1+ {\cal O}(\alpha_s) \right) \,,  
\eea
where the ${\cal O}(\alpha_s)$ corrections can be extracted from~\cite{Gutierrez-Reyes:2018qez}. The $S_{pert.}$ is the perturbative Sudakov factor 
\bea
S_{pert.} = \int_{\mu_b^2}^{s} \frac{d\mu^2}{\mu^2} \left(
A\log\frac{s}{\mu^2} + B \right) \,,
\eea
 with the anomalous dimension $A$ and $B$ can be found in~\cite{Koike:2006fn}.
 The scale $\mu_b = c_0/b_\ast$ where $b_{\ast} = b/\sqrt{1+b^2/b_{max}^2}$  
 with $c_0 \approx 1.22$ and $b_{max} = 1.5\, {\rm GeV}^{-1}$. The $b_{\ast}$-prescription introduces the non-perturbative Sudakov factor $S_{NP}^J$, which can be parameterized as $S^J_{NP} = g_2\ln(b/b_{\ast}) \ln(\sqrt{s}/Q_0)+2 g_hb^2$ with $g_2 = 0.84$, $g_h = 0.042 \, {\rm GeV}^{-2}$ and $Q_0 = 2.4\, {\rm GeV}^2$~\cite{Kang:2014zza}.
 
 The T-odd jet function can be determined by the operator product expansion following the similar strategy in~\cite{Yuan:2009dw}, which we left for future work. The similarity between the T-odd jet function and the Collins function~\cite{Collins:1992kk, Kang:2014zza} suggests the form of its contribution to be
\bea
\partial_{b^\alpha} J_T^q \partial_{b^\beta} {\bar J}^{\bar q}_T
= e^{-S_{pert.}-S_{NP}^T}  \, \frac{b^\alpha b^\beta}{4} \, {\cal N}_q(b) 
{\cal N}_{\bar q}(b) 
\,,
\eea 
 where $S^T_{NP}$ is the non-perturbative Sudakov factor while ${\cal N}_{q({\bar q})}(b)$ is the non-perturbative normalization which determines the sign and the magnitude of the T-odd jet function. Both $S^T_{NP}$ and ${\cal N}_{q({\bar q})}(b)$ could be constrained by experimental analyses. Here for simplicity we parameterize $S^T_{NP} = g_2\ln(b/b_{\ast}) \ln(\sqrt{s}/Q_0)+2 (g_h-g_c)b^2$ with $g_c = 0.0236 \, {\rm GeV}^2$ and the rest are the same as  $S^J_{NP}$~\cite{Kang:2014zza}.

In this example, to further enhance the sensitivity, we again demand the jet charge $Q_J > 0.25$ for one of the jets while $Q_J < -0.25$ for the other. The only change in the factorization is to make the replacement $J^q(b){\bar J}^{\bar q}(b) 
\to J^q(b){\bar J}^{\bar q}(b)
r^i_q r^{\bar i}_{\bar q}$ in $F_{U}$
with $i ({\bar i}) = +\,, -$~\cite{Kang:2020fka}. In $F_T$, we parameterize ${\cal N}_u = - 0.3 r^+_u$ and ${\cal N}_{\bar u} = 0.1 r^+_{\bar u }$ for $Q_J > 0.25$ and ${\cal N}_u =  0.1 r^-_u$ and ${\cal N}_{\bar u} = - 0.3 r^-_{\bar u}$ for $Q_J < - 0.25$. The values of the $r$'s are given previously thanks to the universality. 

\begin{figure}[htbp]
%\vspace{-5.ex}
  \centering
\includegraphics[width=.96\linewidth]{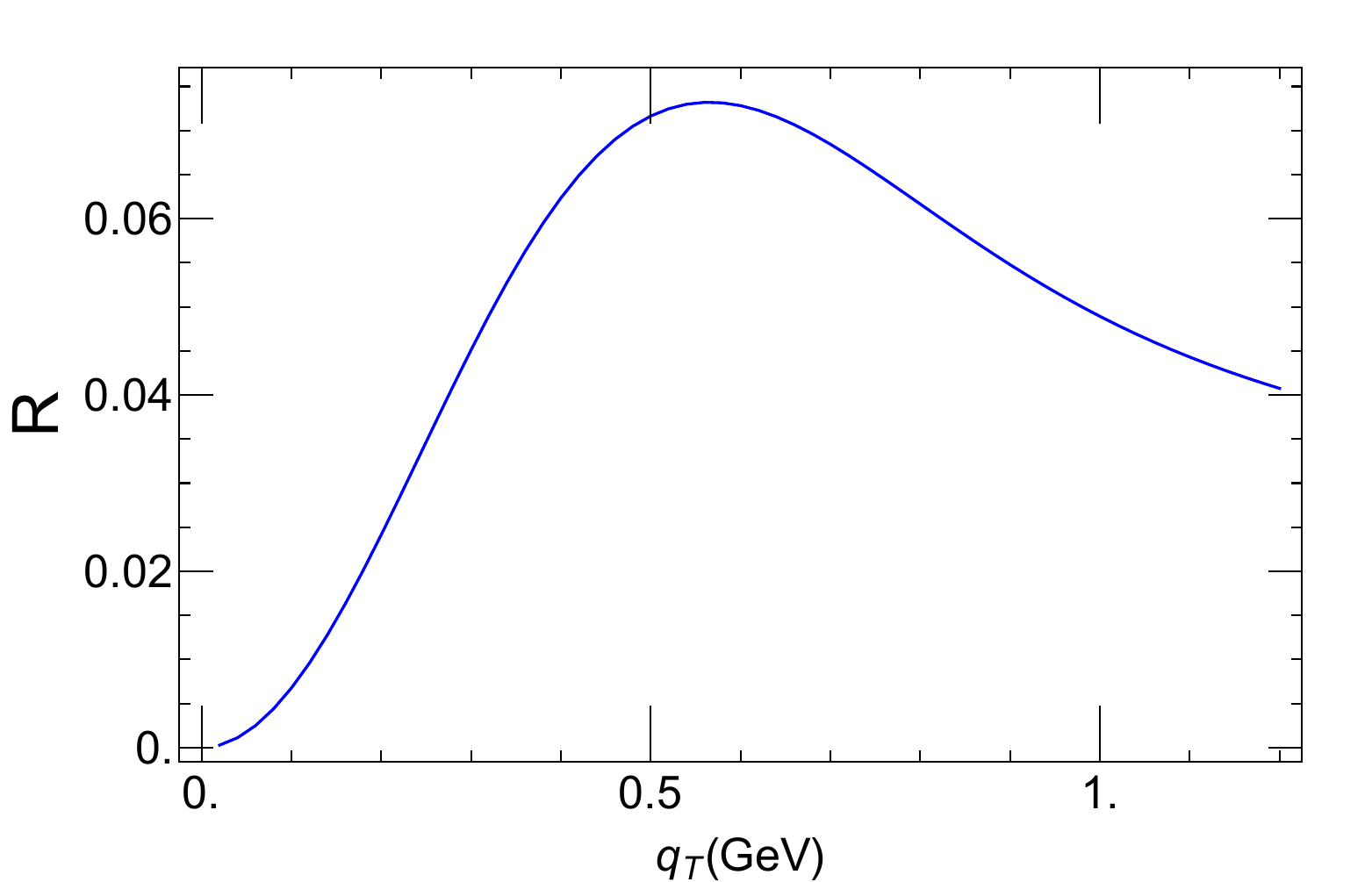} 
 \vspace{0.ex}
\caption{Azimuthal asymmetry induced by the T-odd jets as a function of $q_T$ in $e^+e^-$.}
 \vspace{1.ex}
\label{fig:R}
\end{figure}

In Fig.~\ref{fig:R}, we present a prediction for the azimuthal asymmetry 
$R =  2 \int  d\cos\theta \, \frac{ d\phi_1}{\pi} \cos(2\phi_1)R^{J_1J_2}$ with $\sqrt{s} = \sqrt{110}\, {\rm GeV}$. 
We can see a non-vanishing azimuthal asymmetry induced by the T-odd jet. The actual magnitude and shape of this asymmetry should be determined by upcoming experimental data analyses. Similar azimuthal anisotropy can also be shown to exist in the dijet production in both $pp$ and heavy ion collisions, whose studies %and implications on recent CMS measurements~\cite{CMS:2020ekd} 
will be investigated in the future.

{\it Conclusion.} In this work, we re-examine the factorization theorem of the jet production for nucleon spin studies. Unlike the jet physics at the LHC, for nucleon studies, the focused kinematic region is much lower $\sim {\cal O}(1\, {\rm GeV})$, and thus the jets involved will be inevitably sensitive to the non-perturbative effects. This allows the existence of a T-odd component. We found that the T-odd contribution, which was thought to vanish, gives rise to novel jet phenomena and endows the jets the ability to probe many nucleon spin structures which were thought to be inaccessible by jets. As direct demonstrations, 
we show the proton transversity can couple to the T-odd jet function in the $q_T$ distribution in transversely polarized DIS. We also show the same T-odd jet function can be directly measured via the induced anisotropy in the $e^+e^-$ annihilations at Belle and BaBar. 

\OMIT{Other possible signatures of the T-odd jet contributions could include the azimuthal asymmetries for almost back-to-back jet production in the unpolarized $pp$ collisions. In addition, it could provide an alternative probe of the globally polarized quark gluon plasma in non-central nucleus-nucleus collisions~\cite{Liang:2004ph,STAR:2017ckg}. We left these to future studies.}

We emphasize that the T-odd feature of jets is not limited to the WTA jets discussed currently, but holds for generic jet production and could exist in many jet shape observables in the non-perturbative regime.\OMIT{,due to the correlation with the spin polarization of fragmenting parton.} Furthermore, different jet schemes and/or event shapes weighs differently the non-perturbative soft and collinear contributions, therefore will provide more comprehensive insights into the non-perturbative dynamics. We thus expect the T-odd jets identified in this work to maximize the outreach of the jet probes and\OMIT{bring unique avenues to the spin program at current and future facilities. The T-odd jets could thus} enrich the inputs to the global analyses and complement the conventional hadron probes of the spin structures at current and future facilities.

\OMIT{We emphasize that the T-odd feature of the jet is not limited to the WTA jet discussed in the current work, but holds for generic jet production. In addition, along the similar line, the T-odd asymmetries also exist in many jet shape observables in the non-perturbative regime, due to the correlation with the fragmenting parton spin polarization. The deviation for different jet schemes and/or event shapes are sensitive differently to soft and collinear radiations and weighs differently the non-perturbative interactions from the soft or the collinear sectors, therefore provides us the flexibility to probe differently the non-perturbative mechanisms to the T-odd asymmetries. From this aspect, the T-odd side of a jet, on its own, is extremely interesting to investigate with more details and will shed new insights on the non-perturbative QCD dynamics.}

\OMIT{We thus expect the T-odd ingredient of a jet identified in this work to maximize the outreach of the jet probes and bring unique avenues to the spin program at current and future facilities. The T-odd jets could thus enrich the inputs to the global analyses and complement other probes of the spin information via the conventional hadron process.}

\begin{acknowledgments}
{\it Acknowledgements.} We thank Zhongbo Kang and Yuxiang Zhao for comments on the manuscript. This work is supported by the Guangdong Major Project of Basic and Applied Basic Research No. 2020B0301030008 and by the National Natural Science Foundation of China under Grant No. 12022512, No. 12035007. (H.X.)  and Grant No.~11775023 (X.L.). 
\end{acknowledgments}

%%%%%%%%%%
%\begin{thebibliography}
\bibliography{jet-collins}

\begin{thebibliography}{10}

\bibitem{AbdulKhalek:2021gbh}
R.~Abdul~Khalek {\em et~al.},
\newblock (2021), arXiv:2103.05419.

\bibitem{Anderle:2021wcy}
D.~P. Anderle {\em et~al.},
\newblock (2021), arXiv:2102.09222.

\bibitem{Aschenauer:2016our}
E.-C. Aschenauer {\em et~al.},
\newblock (2016), arXiv:1602.03922.

\bibitem{Adamczyk:2017wld}
STAR, L.~Adamczyk {\em et~al.},
\newblock Phys. Rev. D {\bf 97}, 032004 (2018), arXiv:1708.07080.

\bibitem{Boer:2014lka}
D.~Boer and C.~Pisano,
\newblock Phys. Rev. D {\bf 91}, 074024 (2015), arXiv:1412.5556.

\bibitem{Connors:2017ptx}
M.~Connors, C.~Nattrass, R.~Reed, and S.~Salur,
\newblock Rev. Mod. Phys. {\bf 90}, 025005 (2018), arXiv:1705.01974.

\bibitem{Cao:2020wlm}
S.~Cao and X.-N. Wang,
\newblock Rept. Prog. Phys. {\bf 84}, 024301 (2021), arXiv:2002.04028.

\bibitem{Kang:2011jw}
Z.-B. Kang, A.~Metz, J.-W. Qiu, and J.~Zhou,
\newblock Phys. Rev. D {\bf 84}, 034046 (2011), arXiv:1106.3514.

\bibitem{Liu:2018trl}
X.~Liu, F.~Ringer, W.~Vogelsang, and F.~Yuan,
\newblock Phys. Rev. Lett. {\bf 122}, 192003 (2019), arXiv:1812.08077.

\bibitem{Arratia:2020nxw}
M.~Arratia, Z.-B. Kang, A.~Prokudin, and F.~Ringer,
\newblock (2020), arXiv:2007.07281.

\bibitem{Liu:2020dct}
X.~Liu, F.~Ringer, W.~Vogelsang, and F.~Yuan,
\newblock (2020), arXiv:2007.12866.

\bibitem{Aschenauer:2019uex}
E.-C. Aschenauer, K.~Lee, B.~Page, and F.~Ringer,
\newblock Phys. Rev. D {\bf 101}, 054028 (2020), arXiv:1910.11460.

\bibitem{Arratia:2019vju}
M.~Arratia, Y.~Song, F.~Ringer, and B.~Jacak,
\newblock Phys. Rev. C {\bf 101}, 065204 (2020), arXiv:1912.05931.

\bibitem{Arratia:2020ssx}
M.~Arratia, Y.~Makris, D.~Neill, F.~Ringer, and N.~Sato,
\newblock (2020), arXiv:2006.10751.

\bibitem{Kang:2020xyq}
Z.-B. Kang, K.~Lee, and F.~Zhao,
\newblock Phys. Lett. B {\bf 809}, 135756 (2020), arXiv:2005.02398.

\bibitem{Aschenauer:2017jsk}
E.~Aschenauer {\em et~al.},
\newblock Rept. Prog. Phys. {\bf 82}, 024301 (2019), arXiv:1708.01527.

\bibitem{Zheng:2018ssm}
L.~Zheng, E.~Aschenauer, J.~Lee, B.-W. Xiao, and Z.-B. Yin,
\newblock Phys. Rev. D {\bf 98}, 034011 (2018), arXiv:1805.05290.

\bibitem{Hatta:2020bgy}
Y.~Hatta, B.-W. Xiao, F.~Yuan, and J.~Zhou,
\newblock Phys. Rev. Lett. {\bf 126}, 142001 (2021), arXiv:2010.10774.

\bibitem{Larkoski:2014wba}
A.~J. Larkoski, S.~Marzani, G.~Soyez, and J.~Thaler,
\newblock JHEP {\bf 05}, 146 (2014), arXiv:1402.2657.

\bibitem{Kang:2017glf}
Z.-B. Kang, X.~Liu, F.~Ringer, and H.~Xing,
\newblock JHEP {\bf 11}, 068 (2017), arXiv:1705.08443.
%%CITATION = ARXIV:1705.08443;%%

\bibitem{Kang:2017btw}
Z.-B. Kang, A.~Prokudin, F.~Ringer, and F.~Yuan,
\newblock Phys. Lett. B {\bf 774}, 635 (2017), arXiv:1707.00913.

\bibitem{Chen:2020adz}
H.~Chen, I.~Moult, and H.~X. Zhu,
\newblock Phys. Rev. Lett. {\bf 126}, 112003 (2021), arXiv:2011.02492.

\bibitem{Gutierrez-Reyes:2018qez}
D.~Gutierrez-Reyes, I.~Scimemi, W.~J. Waalewijn, and L.~Zoppi,
\newblock Phys. Rev. Lett. {\bf 121}, 162001 (2018), arXiv:1807.07573.

\bibitem{Gutierrez-Reyes:2019vbx}
D.~Gutierrez-Reyes, I.~Scimemi, W.~J. Waalewijn, and L.~Zoppi,
\newblock JHEP {\bf 10}, 031 (2019), arXiv:1904.04259.

\bibitem{Kang:2020fka}
Z.-B. Kang, X.~Liu, S.~Mantry, and D.~Y. Shao,
\newblock Phys. Rev. Lett. {\bf 125}, 242003 (2020), arXiv:2008.00655.

\bibitem{Ralston:1979ys}
J.~P. Ralston and D.~E. Soper,
\newblock Nucl. Phys. B {\bf 152}, 109 (1979).

\bibitem{Jaffe:1991kp}
R.~L. Jaffe and X.-D. Ji,
\newblock Phys. Rev. Lett. {\bf 67}, 552 (1991).

\bibitem{Jaffe:1991ra}
R.~L. Jaffe and X.-D. Ji,
\newblock Nucl. Phys. B {\bf 375}, 527 (1992).

\bibitem{Cortes:1991ja}
J.~L. Cortes, B.~Pire, and J.~P. Ralston,
\newblock Z. Phys. C {\bf 55}, 409 (1992).

\bibitem{Gamberg:2001qc}
L.~P. Gamberg and G.~R. Goldstein,
\newblock Phys. Rev. Lett. {\bf 87}, 242001 (2001), arXiv:hep-ph/0107176.

\bibitem{qiutalk}
J.~Qiu,
\newblock {Physics at the EIC: QCD at a Fermi scale},
\newblock presented at PHENO 2020, University of Pittsburgh (online), 4-6 May,
  2020.

\bibitem{Brodsky:2002cx}
S.~J. Brodsky, D.~S. Hwang, and I.~Schmidt,
\newblock Phys. Lett. B {\bf 530}, 99 (2002), arXiv:hep-ph/0201296.

\bibitem{Ji:2002aa}
X.-d. Ji and F.~Yuan,
\newblock Phys. Lett. B {\bf 543}, 66 (2002), arXiv:hep-ph/0206057.

\bibitem{Amrath:2005gv}
D.~Amrath, A.~Bacchetta, and A.~Metz,
\newblock Phys. Rev. D {\bf 71}, 114018 (2005), arXiv:hep-ph/0504124.

\bibitem{Collins:1992kk}
J.~C. Collins,
\newblock Nucl. Phys. {\bf B396}, 161 (1993), arXiv:hep-ph/9208213.
%%CITATION = HEP-PH/9208213;%%

\bibitem{Cal:2019gxa}
P.~Cal, D.~Neill, F.~Ringer, and W.~J. Waalewijn,
\newblock JHEP {\bf 04}, 211 (2020), arXiv:1911.06840.

\bibitem{Bauer:2000yr}
C.~W. Bauer, S.~Fleming, D.~Pirjol, and I.~W. Stewart,
\newblock Phys. Rev. {\bf D63}, 114020 (2001), arXiv:hep-ph/0011336.
%%CITATION = HEP-PH/0011336;%%

\bibitem{Bauer:2001ct}
C.~W. Bauer and I.~W. Stewart,
\newblock Phys. Lett. {\bf B516}, 134 (2001), arXiv:hep-ph/0107001.
%%CITATION = HEP-PH/0107001;%%

\bibitem{Bauer:2001yt}
C.~W. Bauer, D.~Pirjol, and I.~W. Stewart,
\newblock Phys. Rev. {\bf D65}, 054022 (2002), arXiv:hep-ph/0109045.
%%CITATION = HEP-PH/0109045;%%

\bibitem{Bauer:2002nz}
C.~W. Bauer, S.~Fleming, D.~Pirjol, I.~Z. Rothstein, and I.~W. Stewart,
\newblock Phys. Rev. {\bf D66}, 014017 (2002), arXiv:hep-ph/0202088.
%%CITATION = HEP-PH/0202088;%%

\bibitem{Bacchetta:2006tn}
A.~Bacchetta {\em et~al.},
\newblock JHEP {\bf 02}, 093 (2007), arXiv:hep-ph/0611265.

\bibitem{Becher:2013iya}
T.~Becher and G.~Bell,
\newblock Phys. Rev. Lett. {\bf 112}, 182002 (2014), arXiv:1312.5327.
%%CITATION = ARXIV:1312.5327;%%

\bibitem{Yuan:2009dw}
F.~Yuan and J.~Zhou,
\newblock Phys. Rev. Lett. {\bf 103}, 052001 (2009), arXiv:0903.4680.

\bibitem{Shanahan:2020zxr}
P.~Shanahan, M.~Wagman, and Y.~Zhao,
\newblock Phys. Rev. D {\bf 102}, 014511 (2020), arXiv:2003.06063.

\bibitem{Zhang:2020dbb}
Lattice Parton, Q.-A. Zhang {\em et~al.},
\newblock Phys. Rev. Lett. {\bf 125}, 192001 (2020), arXiv:2005.14572.

\bibitem{Sjostrand:2007gs}
T.~Sjostrand, S.~Mrenna, and P.~Z. Skands,
\newblock Comput. Phys. Commun. {\bf 178}, 852 (2008), arXiv:0710.3820.

\bibitem{Kerbizi:2021pzn}
A.~Kerbizi and L.~L\"onnblad,
\newblock (2021), arXiv:2105.09730.

\bibitem{Kang:2014zza}
Z.-B. Kang, A.~Prokudin, P.~Sun, and F.~Yuan,
\newblock Phys. Rev. D {\bf 91}, 071501 (2015), arXiv:1410.4877.

\bibitem{Koike:2006fn}
Y.~Koike, J.~Nagashima, and W.~Vogelsang,
\newblock Nucl. Phys. B {\bf 744}, 59 (2006), arXiv:hep-ph/0602188.

\end{thebibliography}

%%%%%%%%%%%%%%%%%%%%%%%%%%%%%%%%%%%%%%%%%%%%%%%%%%%%%%%%%%%%%
\bibliographystyle{h-physrev5}
%\end{thebibliography}

\end{document}